\documentclass[doublecol]{epl2}
\usepackage{epsfig}

\title{Doping evolution of itinerant magnetic excitations in $Fe$-based oxypnictides}
\shorttitle{Evolution of itinerant magnetic excitations in $Fe$-based oxypnictides} 

\author{M.M. Korshunov\inst{1,2} \and I. Eremin\inst{1,3}}
\shortauthor{M.M. Korshunov \etal}

\institute{
  \inst{1} Max-Planck-Institut f\"ur Physik
Komplexer Systeme - 01187 Dresden, Germany\\
  \inst{2} L.V. Kirensky Institute of Physics, Siberian Branch of
Russian Academy of Sciences - 660036 Krasnoyarsk, Russia \\
  \inst{3} Institute f\"ur Mathematische und Theoretische Physik,
TU-Braunschweig - 38106 Braunschweig, Germany}


\abstract{ Employing the four-band tight-binding model we study
theoretically the doping dependence of the spin response in the
normal state of novel Fe-based pnictide superconductors. We show
that the commensurate spin density wave (SDW) transition that arises
due to interband scattering between the hole $\alpha$-pockets and
the electron $\beta$-pockets disappears already at the doping
concentration $x \approx 0.04$ reflecting the evolution of the
Fermi surfaces. Correspondingly, with further increase of the doping
the antiferromagnetic fluctuations are suppressed for $x >
0.1$ and the Im$\chi({\bf Q_{AFM}},\omega)$ becomes nearly temperature
independent. At the same time, we observe that the uniform susceptibility
deviates from the Pauli-like behavior and is increasing with
increasing temperature reflecting the activation processes for the
$\alpha$-Fermi surfaces up to temperatures of about $T=800$K. With
increase of the doping the absolute value of the uniform
susceptibility lowers and its temperature dependence changes. In particular, it is a constant at low
temperatures and then decreases with increasing temperature.
We discuss our results in a context of
recent experimental data.}

\begin{document}

\maketitle

\section{Introduction}

The recent discovery of superconductivity in the iron-based layered
superconductor La(O$_{1-x}$F$_x$FeAs) with $T_c \approx 26$K \cite{kamihara}
has generated the renewed interest in high-temperature superconductivity due to
consequent development of materials with higher T$_c$'s up to $~ 55$K that
contain other rare-earth elements such as Ce, Nd, Sm \cite{chen1,chen2,ren}
instead of La. The physical properties are considered to be highly
two-dimensional; the crystal structure is tetragonal and consists of the LaO
and the FeAs layers which are stacked along the $c$-axis. Similar to many
layered transition metal oxides the superconductivity in oxypnictides occurs
upon introducing doping of either electrons \cite{kamihara,chen1,chen2,ren} or
holes \cite{wen} into the FeAs layers and the parent material shows
antiferromagnetic transition at around 150K
\cite{kamihara,dong,cruz,nomura,klauss}. At the same time, in contrast to
layered cuprates the parent material remains a metal. The observed magnetic
moment per Fe atom has been reported to range between 0.25$\mu_B$ \cite{klauss}
and 0.36$\mu_B$ \cite{cruz} and lies in the $ab$-plane.

There have been various proposals to explain the origin of antiferromagnetism
in these systems. Recent theoretical studies suggest several different
explanations varying from LaOFeAs being an antiferromagnetic semimetal
\cite{cao,ma,yin}, or the system with frustrated magnetic ground state with two
interpenetrating antiferromagnetic square sublattice
\cite{yildirim,fang,ma1,sachdev}. However, the resulting magnetic moments is
larger than that found in experiment thus requiring an inclusion of strong
fluctuations effects that would reduce the magnetic moment. At the same time,
starting from the purely itinerant models it has been also proposed that
LaOFeAs has an antiferromagnetic spin density wave instability due to the
interband nesting of the electron and the hole Fermi surfaces
\cite{dong,klauss,wang}. The resulting magnetic moment has been found to be
about $0.33 \mu_B$ which agrees with experimental data. Despite the right order
of magnitude it remains to be seen whether the strong electronic correlations
that might be important in LaOFeAs  due to the Hund exchange
\cite{haule1,haule2} will modify this result. It has been also argued that a
combined effect of spin-orbit coupling, monoclinic distortions, and $p-d$
hybridization may invalidate the simple Hund coupling scheme \cite{castro}.

In order to understand how the magnetism and the spin fluctuations in
La(O$_{1-x}$F$_x$FeAs) evolve as a function of doping in this letter we present
the study of the magnetic excitations using the tight-binding scheme adopted
previously \cite{korshunov}. In particular, we show that the commensurate spin
density wave (SDW) transition that arises due to interband scattering between
the hole $\alpha$-pockets and the electron $\beta$-pockets at the Fermi surface
(FS) disappears already at the doping concentration $x \approx 0.04$.
Correspondingly, with further increase of the doping the antiferromagnetic
fluctuations are suppressed and at $x \approx 0.1$ the Im$\chi({\bf
Q_{AFM}},\omega) / \omega$ becomes nearly temperature independent. At the same
time, we observe the uniform susceptibility deviates from the Pauli-like
behavior and is increasing with increasing temperature reflecting the
activation processes for the $\alpha$-Fermi surfaces up to temperatures of
about $800$K. With increase of the doping the absolute value of the uniform
susceptibility is decreasing and its temperature dependence changes. It is a
constant at low temperatures and then decreases with increasing temperature.

\section{Theory}
The effective low-energy band structure of the undoped LaOFeAs can be modeled
by the following single-electron model Hamiltonian for the folded Brillouin
Zone (BZ) with two $Fe$-ions per unit cell \cite{korshunov}:
\begin{eqnarray}
H_0 = - \sum\limits_{{\bf k},\alpha ,\sigma } {{\epsilon ^i} n_{{\bf
k} i \sigma } } - \sum\limits_{{\bf k}, i, \sigma}  t_{{\bf k}}^{i}
d_{{\bf k} i \sigma }^\dag d_{{\bf k} i \sigma}, \label{eq:H0}
\end{eqnarray}
where $i=\alpha_1,\alpha_2,\beta_1, \beta_2$ refer to the band indices,
$\epsilon^i$ are the on-site single-electron energies, $t_{{\bf
k}}^{\alpha_{1},\alpha_{2}} = t^{\alpha_{1},\alpha_{2}}_1 \left(\cos k_x+\cos
k_y \right)+ t^{\alpha_{1},\alpha_{2}} _2 \cos k_x \cos k_y$ is the electronic
dispersion that yields hole $\alpha$-pockets centered around the $\Gamma$
point, and $t_{{\bf k}}^{\beta_{1},\beta_{2}} = t^{\beta_{1},\beta_{2}}_1
\left(\cos k_x+\cos k_y \right)+ t^{\beta_{1},\beta_{2}}_2 \cos \frac {k_x}{2}
\cos \frac{k_y}{2}$ is the dispersion that results in the electron
$\beta$-pockets around the $M$ point of the folded BZ. Using the abbreviation
$(\epsilon^i, t_1^i, t_2^i)$ we choose the parameters $(-0.60,0.30,0.24)$ and
$(-0.40,0.20,0.24)$ for the $\alpha_1$ and $\alpha_2$ bands, respectively, and
$(1.70,1.14,0.74)$  and $(1.70,1.14,-0.64)$  for the $\beta_1$ and $\beta_2$
bands, correspondingly (all values are in eV).

\begin{figure}
\begin{center}
\includegraphics[width=0.9\linewidth]{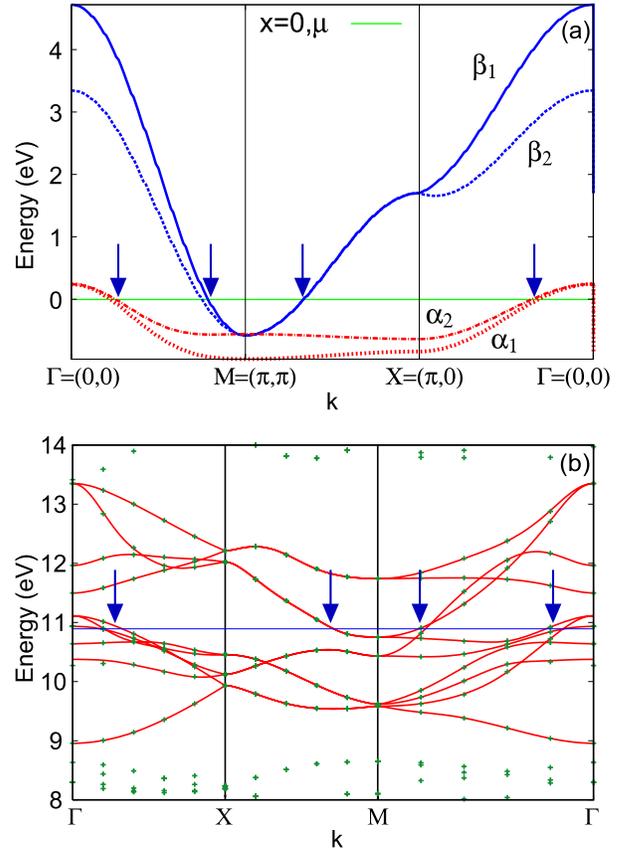}
\end{center}
\caption{(a) Calculated energy dispersion along
the main symmetry points of the first BZ for the undoped, $x=0$, case.
(b) LDA band structure (green crosses) and ten-band model dispersion (red curves)
after K. Kuroki et al. \cite{Kuroki}. Note the difference in BZ directions in (a) and (b).
The large arrows indicate the points where bands cross the Fermi level.}
\label{fig0}
\end{figure}
In Fig.~\ref{fig0}(a) we show the resulting energy dispersion along the main
symmetry directions of the first BZ for the undoped case, $x=0$. The band
structure parameters were chosen to correctly reproduce the LDA Fermi surface
topology and the values of the Fermi velocities for the hole $\alpha$ and the
electron $\beta$ pockets. In particular, we have selected the on-site energies
and the hopping matrix elements assuming the compensated metal at $x=0$ and
calculating the chemical potential self-consistently for the filling factor
$n=4$ (we further assume that there exists another band below the Fermi level
which is fully occupied and not considered here). As a consequence, the hole
Fermi surfaces shifted by vector $(\pi,\pi)$ is nearly completely nested with
that of the electron pockets in full agreement with {\it ab initio} density
functional calculations \cite{dong,haule1,mazin,Kuroki}. Additionally, we take
into account the details of the electronic dispersions of the bands forming the
Fermi surface pockets. In order to visualize the comparison, in
Fig.~\ref{fig0}(b) we present the LDA band structure and the realistic ten-band
model dispersion from Ref.~\cite{Kuroki}. Note, for other doping concentrations
the position of the chemical potential was deduced from the equation $n=4+x$.

\begin{figure}
\begin{center}
\includegraphics[width=0.9\linewidth]{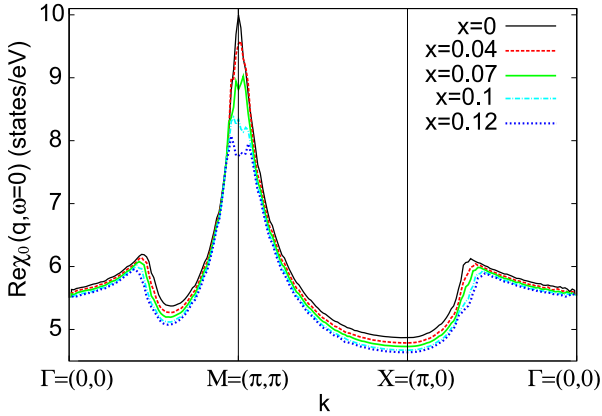}
\end{center}
\caption{Calculated doping dependence of the real part of the physical spin
susceptibility, $\chi_0({\bf q},0) = \sum_{ij}\chi^{ij}_0({\bf q},0)$, where
$i,j$ refer to the band index.} \label{fig1}
\end{figure}

The resulting doping-dependence of the physical susceptibility as obtained by
the sum of all interband and intraband susceptibilities is shown in
Fig.~\ref{fig1}. For the undoped case our results are in qualitative agreement
with that of K. Kuroki et al. \cite{Kuroki} and S. Raghu et al. \cite{raghu},
and with the {\it ab initio} results of J. Dong et al. \cite{dong}.

Within random phase approximation (RPA) the spin response has a matrix form:
\begin{eqnarray}
\hat{\chi}_{RPA}({\bf q},{\rm i}\omega_m)=\left[\mathbf{I}-{\bf
\Gamma} \hat{\chi}_0({\bf q},{\rm i}\omega_m)\right]^{-1}
\hat{\chi}_0({\bf q},{\rm i}\omega_m) \label{eq:chi_RPA}
\end{eqnarray}
where ${\bf I}$ is a unit matrix and $\hat{\chi}_0({\bf q},{\rm
i}\omega_m)$ is $4 \times 4$ matrix formed by the interband and
the intraband bare susceptibilities. For the four-band model considered here the effective interaction
consist of the on-site Hubbard intraband repulsion $U$ and the
Hund's coupling $J$. There is also an interband Hubbard repulsion
$U'$, which, however, does not contribute to the RPA susceptibility. The vertex is given by
\begin{eqnarray}
{\bf \Gamma} = \left[\begin{array}{cccc} U & J/2 & J/2 & J/2
\\ J/2 & U & J/2 & J/2 \\ J/2 & J/2 & U & J/2 \\ J/2 & J/2 & J/2 & U
\end{array}\right].
\label{eq_coupling}
\end{eqnarray}
For the given Fermi surface topology the main magnetic instability in the
folded BZ occurs at the antiferromagnetic wave vector ${\bf Q}_{AFM}=(\pi,\pi)$
due to the interband nesting between the hole $\alpha$- and the electron
$\beta$-bands \cite{dong,korshunov,raghu,mazin}. This is also clearly visible
from our Fig.~\ref{fig1}. Note that in the unfolded BZ  with one $Fe$-ion per
unit cell, the wave vector is ${\bf Q'}_{AFM} = (\pi,0)$ which corresponds to
the `stripe'-like ordering of the $Fe$-spins as observed by neutron scattering
\cite{cruz}. Setting the Hund's coupling to $J = 70$meV and choosing $U=320$meV
we obtain the ordering temperature $T_N = 138$K as determined by $\det
\left[\mathbf{I}-{\bf \Gamma} \hat{\chi}_0({\bf q},{\rm i}\omega_m)\right] =
0$. Solving the condition for the SDW instability below T$_N$ which can be
regarded as a mean-field equation for the SDW order parameter, we obtain
$\Delta_{SDW}(T=0K) = 31$meV which corresponds to the magnetic moment per two
Fe sites to be $\mu \approx 0.33 \mu_B$.
\begin{figure}
\begin{center}
\includegraphics[width=0.9\linewidth]{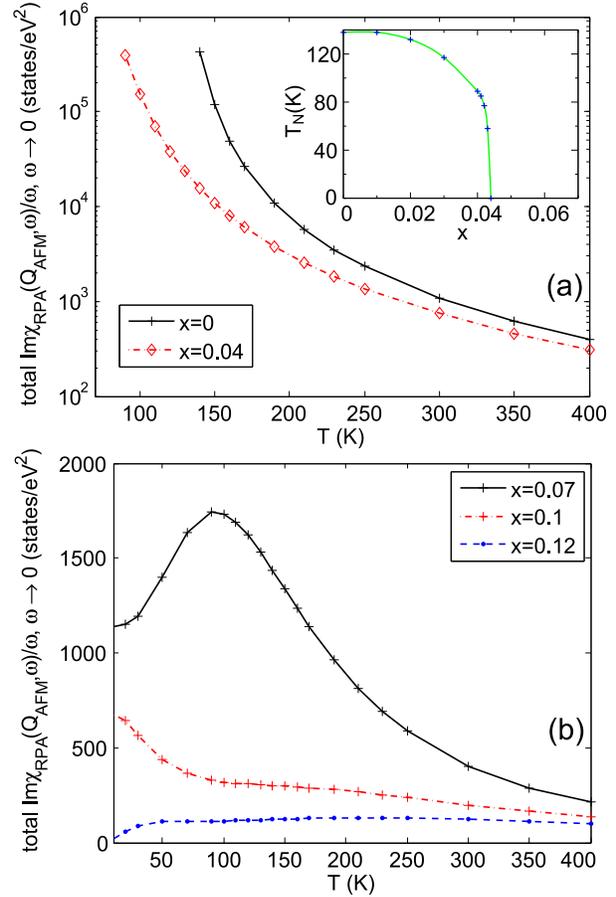}
\end{center}
\caption{Calculated $\lim_{\omega \to 0}$ Im$\chi_{RPA}({\bf Q}_{AFM},
\omega)/\omega$ for various doping concentrations. Note the $\log$ scale in
(a). The inset in Fig.2(a) shows the calculated doping dependence of the
N\'{e}el transition temperature.} \label{fig2}
\end{figure}

Note, the small values of $U$ and $J$ used here is a consequence of the absence
of the self-energy corrections within RPA approach. Such corrections would
reduce the value of the absolute magnitude of the spin susceptibility and
correspondingly yield larger values of the coupling constants $U$ and $J$.

In the inset of Fig.~\ref{fig2}(a) we show the doping dependence of the
N\'{e}el temperature. One finds that it decreases quite rapidly as a function
of doping and goes to zero already at $x \approx 0.04$. The reason of the rapid
suppression of the N\'{e}el temperature is quite obvious. Away from $x=0$ the
spectral weight of the hole $\alpha$-pockets at the Fermi surface is decreasing
and the condition for interband nesting becomes worse as it is readily seen
from Fig.~\ref{fig1}. In particular, one finds that the peak at the
antiferromagnetic wave vector, {\bf Q}$_{AFM}$, decreases quite rapidly away
from $x=0$. Remarkable that such a small deviation from undoped case changes
the situation dramatically also in the NMR experimental data \cite{nakai} that
gives an additional support in favor of the nesting scenario of the
antiferromagnetic transition.
\begin{figure}
\begin{center}
\includegraphics[width=0.9\linewidth]{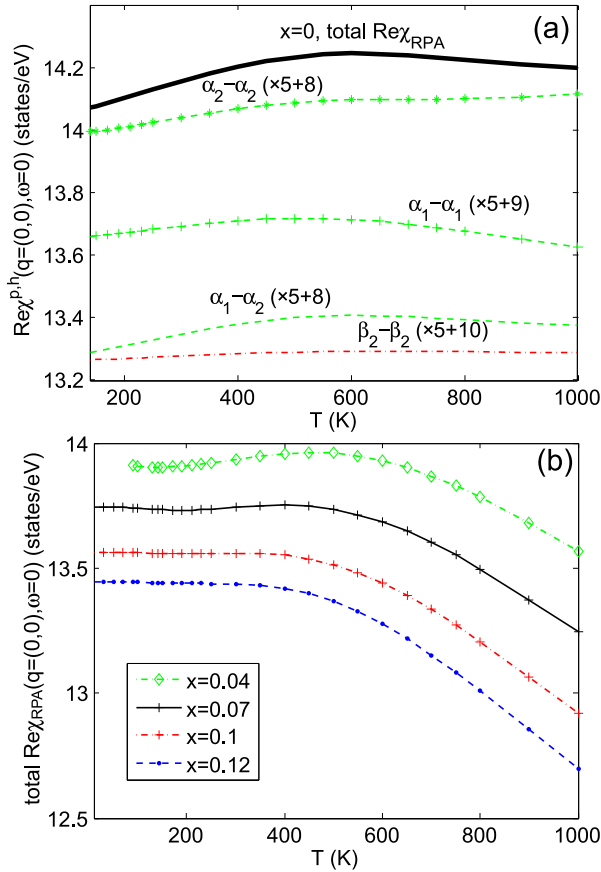}
\end{center}
\caption{Calculated uniform part of the
total spin susceptibility, $\chi_{RPA}({\bf q}\to 0, \omega \to 0 )$
as a function of temperature at $x$=0 (a) and for other doping
concentrations (b). In (a) we show also the partial contributions
including the interband and the intraband transitions.}
\label{fig3}
\end{figure}

In Fig.~\ref{fig2} we show the doping dependent evolution of the
$\lim_{\omega \to 0}$ Im$\chi({\bf Q}_{AFM}, \omega)/\omega$. One
finds that at $x=0$ it diverges at T$_N$ and with further increase
of the doping the antiferromagnetic fluctuations are quickly
suppressed. Remarkable one finds that Im$\chi({\bf Q}_{AFM},
\omega)/\omega$ at $x=0.12$ does not show any enhancement
characteristic for strong antiferromagnetic spin fluctuations and
stays nearly constant. Despite the fact that the real part still
shows the peaks around ${\bf Q}_{AFM}$ the antiferromagnetic
fluctuations are quite strongly suppressed in the imaginary part of
the spin susceptibility. This is due to the fact that the RPA
response has a matrix form and thus the damping of the fluctuations
is quite strong. At the same time, one has to keep in mind that the
spin dynamics on the As sites originating from the stripe-like
ordering of the Fe spins as probed by NMR will be suppressed due to
hyperfine interaction. Therefore, further experimental studies are
necessary to understand the evolution of the antiferromagnetic
fluctuations in these systems.

In Fig.~\ref{fig3} we show the temperature dependence of the uniform
susceptibility for various doping concentrations. It is interesting to note
that the uniform susceptibility above T$_N$ does not show either the Pauli-like
or the Curie-Weiss type behavior. At zero doping concentration the total
susceptibility is increasing as a function of temperature up to 600K and then
decreases following the Curie-Weiss like behavior. Looking on the partial
contributions, one finds that this temperature dependence is determined mainly
by the transitions between $\alpha$-bands which produce the hole pockets around
the $\Gamma$-point in the BZ. In particular, the hole-like Fermi surfaces of
the $\alpha_1$ and $\alpha_2$ bands are only slightly splitted. The gap between
the two Fermi energies that would occur for zero {\it transferred} momentum is
about 50meV. Therefore, due to the temperature activated transitions between
two $\alpha$ bands, the interband susceptibility will increase with increasing
temperature up to 600K and then decrease. This slight increase of
susceptibility is quantitatively consistent with available experimental
measurements \cite{klauss,klaussprivate}. Note that the transitions within
$\beta$ bands and between $\alpha$ and $\beta$ bands show almost Pauli-like
behavior. Upon changing doping the overall magnitude of the susceptibility is
decreasing which reflects the reduction of the total susceptibility as also
shown in Fig.~\ref{fig1}. The latter occurs due to the filling of the hole
pockets. We also observe the change in the temperature dependence of the
uniform susceptibility. For $x>0.1$ the uniform susceptibility is constant up
to 200K and then decreases with increasing temperature. This change occurs due
to the filling of the $\alpha$-bands upon varying doping and the reduction of
their relative splitting as can be seen from Fig.~\ref{fig0}.

We finally note that in our analysis we assume all matrix elements for
calculations of the spin susceptibility to be unity. Note that a qualitative
agreement between our results and those found in Refs.~\cite{dong,raghu,Kuroki}
for the undoped case seems to justify our approach. In addition we further
neglect the other three-dimensional band that gets quickly filled upon doping.
Although its inclusion may be important with regard to the formation of
three-dimensional N\'{e}el order, it will not change much the doping dependence
of the two-dimensional in-plane magnetic fluctuations.

\section{Conclusion}
We have analyzed the doping dependence of the spin excitations in
La(O$_{1-x}$F$_x$)FeAs based on a purely itinerant model. We find that the
interband antiferromagnetic spin fluctuations are rather rapidly suppressed and
the N\'{e}el temperature disappears already for $x \approx 0.04$. With further
increase of the doping the short-range antiferromagnetic fluctuations disappear
at $x\approx 0.12$ in agreement with NMR data. Given the fact that the
superconductivity seems to be strongest at this doping concentration it is
interesting to see whether these fluctuations can be responsible for the
formation of superconductivity. Due to the multi-orbital character the uniform
susceptibility shows neither Pauli-like neither Curie-like behavior. In
particular, for small doping the total susceptibility is increasing up to 600K
and then decreases. With increasing doping the susceptibility stays constant at
small temperatures and then lowers with increasing temperatures. We find that
this characteristic behavior originates from the interband transitions between
slightly splitted $\alpha$-bands.

{\it Note added:} After submission of this manuscript we became aware of the
study by Anisimov et al. \cite{Anisimov}, where the values of the average
Coulomb repulsion $U$ and Hund's exchange $J$ were obtained by the first
principles constrained density functional theory in Wannier functions
formalism. Due to the delocalization of Wannier functions for the $Fe-3d$ basis
set, the Coulomb parameters were significantly reduced in comparison to their
intraatomic values and became $U \approx 0.6$eV and $J \approx 0.5$eV. These
are close to the effective parameters used in the present study.

\acknowledgments We would like to thank  B. B\"uchner, S.-L.
Drechsler, D. Efremov, P. Fulde, I. Mazin, R. Moessner, and D. Parker for useful
discussions. I.E. acknowledges support from Volkswagen Foundation.

\end{document}